\documentclass[12pt]{article}
\usepackage{amsmath,amssymb,latexsym}
\usepackage{graphicx}

\begin{document}

\Large
\begin{center} 
The Galactic Dynamo and   Superbubbles\\
\vskip .2 in
Russell M Kulsrud\\
\end{center}
 
\normalsize

\section{Abstract}

In previous galactic dynamo theories of the origin  
of the magnetic field in our galaxy, the
subject of flux-freezing has been omitted.
As a consequence, the equation of mass flow
has generally also been omitted, particularly
in the halo where the galactic gravitational field
will operate on the mass flow.  In this paper it has been shown
that this  neglect could have serious consequences for
the results obtained from those galactic dynamo simulations
that include the halo.  A modification of 
these  dynamo theories is proposed which involves the
expulsion of very small bits of the magnetic field lines,
rather than the  wholesale expulsion of the  complete magnetic lines
 encapsulated in the previous theories.
This expulsion is accomplished by a spike instability 
that arises from superbubbles, when they break
out of the galactic disc and their shell fragment.
This leads to a {\it cut } in the lines of force
that still remain in the disc.  Subsequently, 
normal disc turbulence rotates the {\it cut} lines
and thus dissipates their mean flux
removing them from a role in the dynamo theory. 
This new process takes a length of time comparable,
but a little longer than the previous growth time
of the disc dynamo, but avoids the previous difficulties
associated with flux-freezing and flux expulsion. 

\section{Introduction}

In previous galactic dynamo simulations, 
the mean-field equations have been  solved for the
magnetic field, but the equations for the 
mass flow have been  neglected.  This is not important
for the solutions in the galactic disc, because there the  mass 
flows are very slow.  However,  it is important
in the halo in which the magnetic flux is diffused
upward across the halo to entirely escape from the galaxy. 
  This escape is necessary for our galaxy  because the  
total   flux must be  conserved and the 
flux in any dynamo origin 
 theory    must
 start from a very small value and grow to its present
value. In our galaxy the present flux  is observed to be non zero.

 However,  during the amplification the mass which 
is tied to the flux by flux-freezing must also 
diffuse out of the galaxy with the magnetic flux.
This mass  is acted on by
the gravitational field of the galaxy. This 
gravitational  field can pull the mass and flux tied
to it back into the galactic disc 
 as fast as the halo turbulence is diffusing the flux
upward.
Indeed, a simple estimate shows that
for  the generally accepted value of the   diffusion coefficient  
 (of order of $ 10^{ 28} \mbox{ cm}^2/s $ )
the gravitational force dominates over the diffusion
and the dynamo is   strongly inhibited. This estimate comes
from  an elementary one-dimensional  problem that
includes diffusion, flux-freezing, and gravity.

To get around this difficulty 
a new idea is presented in this paper for a process that 
dissipates the flux by 
 the  amount  required by
any dynamo origin  theory.  This process does not
require the removal of  
 mass from the galaxy, as is required by previous origin  theories
when flux-freezing is included. 
It  involves  those galactic  superbubbles that break out of the
 disc.  It is known that  when a  superbubble shell
does  break out of the galaxy its    shells break up into many fragments,
and from each of these fragments,  a new instability, the {\it spike }
instability    can
develop.   This instability    accelerates  a    small mass,
and the  small piece of the  flux 
swept  up by  it, 
to faster   than  the galactic escape velocity allowing  
 these  pieces to  ballistically 
escape  from the galaxy.
As a  consequence,     the part of the flux line  
attached to it remaining in the disc is   effectively
cut into short pieces of finite length.  These
pieces,  although
 connected to each other through the expelled pieces,
 are not inhibited
from rotating freely and in this way  way {\it flux}
 of these lines  is  dissipated.
Although very little mass is actually removed from 
the galaxy and the rest of the mass on the the line
remains in the disc, its flux not longer counts
in the balance of flux in the dynamo theory. 
It  turns out that this {\it in sit tu}  dissipation is 
equivalent to that
resulting from flux expulsion,  but it not subject to
the gravitational difficulty arising from the 
 flux-freezing constraint.  

This paper is devoted to elaborating how this second process
can enable the alpha-omega dynamo to function without
the difficulties involved in flux and mass  escape.

In the next   section, an attempt is made  to quantify 
the difficulties encountered
when flux-freezing and the effect of gravity on the
related mass are  included in the dynamo simulations.
These are semi quantified by a simple one-dimensional model, and shown
to be potentially serious.

In the third     section, I present  the proposed new solution, and 
introduce   the spike instability and the relevant   properties of the superbubbles.
  In sections four and five,  I describe the superbubble phenomena
and its properties which underlie the process. In  section six,  I derive 
the linear and non linear properties of the 
spike instability.   In section seven, I give an estimate
for  the number of spikes, the rate at which  any relevant 
line is cut  per unit time,  and the mean length of the cuts.
In  section eight   I calculate  the rate of rotation of the relevant
cut lines of force, which depend on their length,
 and  the time for  their flux to be dissipated by this rotation.
This time is the important time which sets a limit on
the speed of the dynamo if no line escape is possible.   
The number of lines
whose lengths suffer the spike instability is also estimated.

The tentative conclusion from this: in approximately a  billion years
 enough cuts are produced that  every line in the outer
quarters of the disc, loses its mean flux.  This implies
that the dynamo can grow the galactic field  without the need for 
actual flux escape.
However, the rate of dynamo growth  is somewhat slower than
thought at first.

\section{ The inhibition of previous  galactic dynamos by flux-freezing}

   Our galaxy has a strong large-scale magnetic field
and its  origin is  of  considerable astrophysical interest.
Where does it come from?  It is generally agreed that
the galaxy  started off with a very weak  seed  field, perhaps
as weak as $ 10^{ -16} $ gauss produced by cosmological
processes, (Kulsrud et. al. 1997 ; Moss and Sokoloff 2013). 
This field is  believed to be increased exponentially  
by  a galactic dynamo. The problem is complicated for our
galaxy, which is known to have a non zero flux frin 
 measurements which indicate that the field does not change
sign across the galactic midplane.  This implies that some
flux must leave the galaxy.  

A large number of important papers have been written on this galactic dynamo.
Examples of these 
 include the following references:
  Baryshnikova et. al. 1987;
Brandenburg et al 1992;  Brandenburg et. al. 1993;
Dobler et. al. 1996; Donner and Brandenburg: 1990; Moss and Brandenburg 1992;
 Moss, et. al. 1993;  Poezd et. al. 1977, Ruzmaikin et. al. 1988
Beck et. al. 1996; and  Parker 1997.
In these papers the structure of the galaxy is taken as a  thin disc
with thickness $ 2 H \approx 400 -1000 \mbox{ pc} $.
This disc is embedded in a large sphere of radius
$ 4-10 \mbox{ kpc} $. The region of the galaxy outside 
of the disc is called the halo.

In these papers the mean-field dynamo 
equations (Steenbeck et. al. 1966) are successfully 
solved throughout the sphere with solutions
that grow exponentially in a growth time of a few hundred
million years.

  These equations are 
\begin{eqnarray} \label{eq:A} 
\frac{\partial B_r}{\partial t} &=&
-\frac{\partial (\alpha B_{\theta})}{\partial z} + 
\frac{\partial  } {\partial z}  \left(  \beta 
\frac{\partial B_r}{\partial z} \right) \\ \nonumber 
 \frac{\partial B_{\theta}}{\partial t} &=&
 r\frac{\partial \Omega }{\partial r} B_r + 
\frac{\partial  } {\partial z}  \left(  \beta 
\frac{\partial B_{\theta }}{\partial z} \right) 
\end{eqnarray} 

In these equations alpha, which produces the famous
alpha effect,  turns toroidal flux into poloidal flux.
$ \Omega(r)$ is the differential rotation of the galaxy,
and it  turns poloidal flux into toroidal flux.
$ \beta $  is the coefficient of turbulent magnetic diffusion 
which mixes up the magnetic fields. 
To allow growth of the galactic field, which starts with
zero flux early on and at present has  non zero flux, conditions
must be satisfied on the surface of  the halo that allow the escape
of flux into the intergalactic medium.

These integrations are mathematically correct, but they ignore
two important  physical effects, which   throws some doubt on the physical
soundness of their results.  

These two effects are:

1. Flux-freezing in detail.  The mean  field equations  given
in Equation~(\ref{eq:A})  do preserve flux-freezing on the large
scale, but they do not take into account
that  even on small scales any mass tied to a given flux tube must remain
tied to  this flux tube as it moves.  By ignoring this property
the solutions  have been able to avoid
 the dynamics of the mass flow and are solved without regard to
the mass dynamics, which would follow from flux-freezing.

2.  The  gravitational field.  Once mass flow is
ignored there is no need  to take into account the gravitational 
field  of the galaxy.  However,  it has a strong effect on any  mass flow
tied to the flux flow.
In any simulation in which  the mass is tied to the flux,
 the gravitational field plays an important role. 

In order to understand   the importance of these effects
most clearly, I consider only the earliest phases of
the galactic dynamo.  This is the phase in which
 when the Lorentz forces are too  weak
to be macroscopically important, but strong enough 
to enforce flux-freezing.

 I show by a simplified model, that when these two 
effects are included in the simulation, the gravitational
field will strongly affect the flow
of the flux  out of the halo.  To this purpose,
I ignore all dynamo terms except the beta terms, and I assume
that $ \beta $ is constant.

 The flow of the flux  is treated most simply by considering
what happens to a bundle of flux which   leaves  the disc
at time $ t=0 $.  I  assume one space dimension,
 $ z $, in the vertical direction 
normal to the disc.  Let $ f $ be the intensity of the bundle
and let us  describe  it  by  a collection of particles diffusing with 
diffusion coefficient $ \beta $, and under the influence
of  a downward gravitational 
field $ g $. It is physically reasonable to assume that
the diffusion will occur in an downward accelerating
frame with a diffusion  coefficient whose value is independent of the 
acceleration. The frame is accelerating downward  
with at a constant
value $ g $.

Then as shown in the appendix,  $ \Phi(t) $ 
 the fraction of this bundle  of particles
that is beyond 
the radius of the galaxy, $ R $ at time $ t $ is,  
\begin{equation}\label{eq:BB} 
\Phi(t) = \mbox{Erfc} \left[ (R+g t^2/2) /\sqrt{4 \beta t} \right] 
\end{equation}
where Erfc is the compliment of the error function.

If $ g=0 $ the simplified model corresponds to the previous simulations
	     and the  fraction  escaping is simply
\begin{equation}
\Phi(t) =\mbox{ Erfc}  \left[R /\sqrt{4 \beta t} \right] 
\end{equation}

As $ t $ increases, the argument of the complimentary error
functions decreases, and $ \Phi $ increases.  Finally, when
when $ t \gg t_c $ where
\begin{equation}
t_c=\frac{R^2}{4 \beta}  =8 \times 10^7 R_{10} /\beta_{28} \mbox{ years}
\end{equation} 
 the argument of $ Erfc $ is much less than unity , and the bulk
of the flux is  beyond $ R $. $ R_{10} $ is the radius
of the halo in units of 10 kpc and $ \beta_{28}  $ is 
$ \beta $ in units of $ 10^{28} \mbox{ cm}^2/s $.
These  values are  consistent with the dynamo times
times attained in the dynamo models, which shows  that the example
gives roughly the correct results. 

If we include the galactic gravitational field,
 $ g \sim 10^{ -8} \mbox{ cm}^2/s $, then the fraction
of lost particles (or flux) is given by the full 
Equation~(\ref{eq:BB}).  In this case,  as $ t $ increases,
 the argument of Erfc at first decreases till it reaches
a minimum value at $ t=t_g $, where 
\begin{equation}
t_g =\sqrt{\frac{2 R}{3g}}
\end{equation} 
and then increases.

That is, the lost flux in our model increases to
$ \Phi(t_g) $, and then seems to decrease.  In our model
this corresponds to particles that have crossed $ R $
and then  fallen  back inside $ z=R $.  Presumably,
conditions beyond $ z= R $ are different. For example,
$ g $ may weaker and other factors may enter not included
in the model.  Therefore, assuming any flux that has passed 
beyond  the halo
boundary  escapes, the largest fraction of the  flux that can escape
  is
\begin{equation}
 \Phi_g(t_g) = \mbox{ Erfc} [\frac{(2 R/3)^{3/4} g^{1/4}}{\beta^{1/2}}]
\end{equation}
$ t_g $ is essentially the free fall time from the radius $ R $.
Numerically, we have the following table.

 For $ R = 10 \mbox{ kpc} $ and $ \beta = 10^{ 28} \mbox{cm}^/s,   
\Phi =10^{-13.9} $. 

 For $ R = 5 \mbox{ kpc} $ and $ \beta = 10^{ 28} \mbox{cm}^/s,  
\Phi =10^{-5.33} $. 
 
 For $ R = 10 \mbox{ kpc} $ and $ \beta = 10^{ 29} \mbox{cm}^/s,
\Phi =0.016 $. 

 For $ R = 5 \mbox{ kpc} $ and $ \beta = 10^{ 29} \mbox{cm}^/s, 
\Phi =0.191 $. 

Thus, we see that if  $ \beta $ is of order $ 10^{ 28} \mbox{cm}^2/s $ 
(Poezd 1993, Dickey and Lockman 1990, Kulkarni and Fich 1985)
the fraction  of the  flux that escapes is extremely small. 
and  that  nearly all of the flux escaping from the {\it disc}
 is reflected
back by  the halo.  That is to say 
 the disc boundary conditions are essentially
closed  conditions, significantly  modifying the standard 
dynamo solution.

This simple  approximate model implies that at the very least
the standard simulations
should  be repeated  with mass attached to the flux.  by flux-freezing,  and 
a gravitational field included.   This should be the case
 even when  one  has a   larger diffusion 
coefficient. 

by the way, if the diffusion coefficient  were large enough,
(i. e. $ 10^{ 29} \mbox{ cm}^2/s ) $  to produce 
the escape of the flux in the presence of flux-freezing
and gravity,  a considerable amount of energy is involved.
In fact, if a sufficient  amount of flux escapes to allow the 
dynamo to double the magnetic flux in  the disc,
at least half of the disc will be lifted out of the
galaxy.   This will take approximately $ 10^{ 55 } $ ergs
for an annulus with width a kpc.
If the removal occurs over a time of a billion years,
the amount of power that must be supplied is of order
$ 10^{ 39} $ ergs per second.   This is and interesting
number.  The power need to resupply
the loss of cosmic rays is a little larger.  About
$ t $  times $ 5 \times 10^{ 40} $  ergs per second and infalling mass
has  been invoked to supply this energy.
   
Clearly, the above argument rests heavily on the property
of flux-freezing.  To see that flux-freezing is valid in this
case,  consider the magnetic diffusion equation, 
in a static medium, with non turbulent 
resistivity, 
\begin{equation}
\frac{\partial \delta B}{\partial  t} =
\frac{\eta }{4 \pi } \nabla^2 B
\end{equation}
where according to Spitzer (1962) $ \eta/4 \pi = 10^{ 7}/T_e^{3/2} 
\mbox{ cm}^2/s $
where $ T_e $ is the electron temperature in electron volts.
Let $ B $ have a scale length of $ \ell $ and estimate 
$ \nabla^2  B \sim  B/ \ell^2 $.  Then with $ T_e = 1 \mbox{eV} $
(a typical astrophysical temperature) ,
\begin{equation}
\frac{\delta B}{B} \sim \frac{10^{ 7} t}{\ell^2}
\sim \frac{3 \times 10^{ 23}}{\ell^2} = 3 \times 10^{ -14}
\end{equation}
if $ t = 10^{ 9 } \mbox{ y } $ and $ \ell \sim \mbox{ 1 pc} $.
If the plasma is moving, then this change in $ B $ represents
the slippage of magnetic flux relative to the plasma.
For the galactic dynamo these are appropriate values for
$ t $ and $ \ell $ as the flux leaves the disc, and 
this very small slippage justifies the flux-freezing assumption.
In any event, if $ \beta $ is not sufficiently large, then
either  the standard galactic  dynamo and needs to be 
revisited or flux-freezing must be invalid.  

In light of this, it would seem prudent to look for a 
different way to get rid of the flux,  and it is the
purpose of this paper to suggest such an alternative.

\section{ Alternative suggestions for a solution  to the problem}

Clearly,  if the halo diffusion coefficient is not large enough
something more is needed  to make the galactic dynamo  viable.
  Parker (see  figure 3c in
Parker 1992) has suggested
that it is not necessary for entire lines to  be bodily moved out of the halo,
but only partial loops of the lines.   The mass on the loops themselves
could be greatly reduced by mass sliding down the
field lines, and the loops  lifted upward by
buoyancy or  cosmic ray  pressure.
 Parker  then suggests  that these loops would
be cut loose by magnetic reconnection and lifted out
of the galaxy  by buoyancy and cosmic ray  forces.
However,   Parker's lifting  mechanism is not available
in the  early very weak field phase of the galactic disc.
It is also true that in   Parker's model most of each   flux line would be left
behind and so the disc flux would  appear to be unchanged.
However,  in section 9, I show that, even without reconnection. 
 such pieces of the field lines
could be dissipated by horizontal turbulent diffusion.

Other plausible  approaches to the flux removal problem
are the  use of  supernovae or superbubbles to  lift portions
of the field lines out of the galaxy.   In  these approaches,
  the main flux is also left behind in the disc,  although these
remnants of the field lines might also be dissipated as in section 9.   
However,   neither 
cosmic rays, supernovae or superbubbles  have  the capacity
to directly lift even pieces of enough  lines  out of the galaxy.
Most type II supernovae   are too embedded
in the disc to  plow their way through the interstellar 
medium and leave the disc. Only the very few type II  supernova
that explode high above the disc could  throw material
out of the halo, and there are  not enough of these.
 The largest  superbubbles
 potentially have enough power, but once their shells  leave the disc
they become unstable and break up into fragments which fall
back into the disc (Rafikov and Kulsrud 2000, Mac Low et. al.  1989).

This is true of our galaxy,  but not necessarily true
of others.  In fact, in starburst  galaxies
the sources of activities are much stronger and strong plasma
outflows from then are observed. But our galaxy 
is not forming stars at the rates of these galaxies,
and there is no evidence that it ever did. On the other hand, smaller
galaxies have weaker gravitational fields that 
could  allow outflows.   

Thus, the possible solutions to the   flux problem in our early galaxy 
  face  {\it two}  difficulties: (1) First, the existence of
enough sources sufficiently powerful to 
expel mass from the halo. (2) Second,  the removal
of the net flux of any piece  of a line left  behind.

In this paper  I suggest  a different approach to the problem:

  First, I propose that fragments of  a superbubble shell
which breaks out of the disc
 are unstable to forming spike-like structures
that can drive small lumps of mass upward.  The mass in  these
{\it spikes} acts like the  mass in Parker's loops,  in that, mass can slide
down their sides  reducing the mass on their tops.  The
 mass  remaining on their  tops
  can then be accelerated  to   the galactic escape velocity, 
and  escape ballistically   out of the galaxy.
 This mass
will carry    a small portion of the lines of force
   out of the halo. This will resolve the first difficulty.

Second, it is true
that the  line removal leaves the rest of the line
in the  disc,  but    this remainder is  effectively {\it cut}
 into separate  pieces.  These pieces will be 
rotated by the turbulent motions  in
 the disc. These pieces are
still connected to each other through the  pieces of flux
expelled by the spikes.  However, it easily seen that the flux
in the disc is thus reduced, the residual  part of the flux being carried away 
by the tiny pieces of expelled flux. Thus,  surprisingly,
 the flux will be  reduced by 
the cuts due to   the spikes, without
 the bulk of the flux lines or the  mass being removed from the disc.  

 If there were no spikes
 the lines of force could abstractly still be considered to be divided into
pieces much as they are with the spikes. Because the 
magnetic field is very weak the pieces could still rotate
randomly. But 
in this case,  these pieces are connected to each other through
 flux lines still in the disc so that no flux
is lost. 

The two cases are  illustrated in figure  1.  
On the left we have a line of force AD in the xy plane.
Imagine it divided into two parts  AB and CD.  Let the two parts be
rotated, as in the figure,  to A'B' and C'D'.  These two parts have
no flux in the disc  in either the $ x $ direction or the $ y $ direction .
However  they were connected by a short line BC
which by flux-freezing  goes to B'C',  so that the combined line A'B'C'D' has the
same flux in the disc as the original line.  

Now, consider the same situation and the same line segments, except
B'C' has been replaced by a line that stretches out of the disc
to the spike and back.  This segment has very little flux in the disc
and the remaining lines have no flux. As far as the disc is concerned
the line has been {\it cut }  into two parts whose disc fluxes can cancel.

It should be noted that if the line
$ B'C' $ reconnected with itself,  then the
matter along it   would naturally drop back into 
the disc and draw the line $ B'C' $ with it.
Thus, reconnection would destroy the model.
But it seems highly unlikely that flux-freezing
on this scale would allow reconnection.  

\begin{figure}
\rotatebox{0}{ \scalebox{0.55}{ \includegraphics{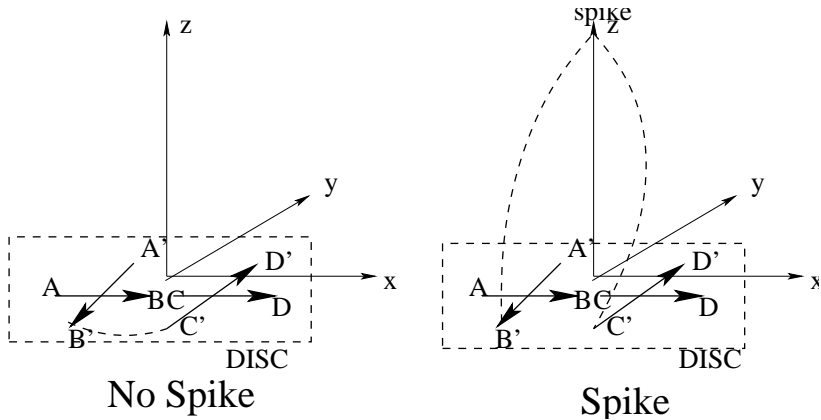} } }
\caption{The line breaking with and without the spike}
\label{Fig1}
\end{figure}
\vspace*{1in}
 
As the cut pieces rotate and decrease the disc  flux,
the magnetic line in the expelled parts  will
twist in the opposite direction  consistent with
flux and helicity is conservation. However, the helicity in the
disc, under any  plausible definition of local helicity, 
will change,  as required for any dynamo that is growing the
field. 

In the rest of this paper I show that in a billion years,
there are enough spikes produced
by superbubbles  to cut enough lines
to reduce the flux  by a finite value in the top and bottom
parts  of the disc.  This will resolve  the second difficulty.

\section{SuperBubbles}

 Next, we discuss the superbubbles.

   Superbubbles were first discovered by Heiles in 1974.
(See Heiles 1974)
He found them by surveying the galactic disc in narrow, 10 $ km/s $, width
Doppler-shifted velocity bands of the
21 centimeter line. 
He found that in each band,  the  intensities were  located in
definite stripes.  Combining these stripes in velocity and space, he 
 interpreted them as gigantic expanding bubbles.
These bubbles were similar to those produced by stellar winds,
(Weaver et al 1977), but on a much greater scale.

It was soon  realized that  superbubbles
must be produced by multiple sequential supernovae
originating in a  forming galactic cluster of stars.
The O and B stars in the cluster  explode as supernovae almost
as soon as they form,  so that their energy instead
of going  into the  kinetic energy of their remnants,
 goes   into 
a bubble  of high-pressure low-density plasma.
This plasma   pushes all
the surrounding mass into a massive shell.
An example of the larger of these  superbubbles is  the North Polar Spur.
It has  an enormous  diameter  of more than
80 degrees stretching across the whole sky.
  After a period of   time, it was appreciated  that nearly every
superbubble could be identified with a young galactic cluster,
and conversely every galactic cluster was generating a superbubble,
(Elmegreen and Clemens 1985).

In these superbubbles the supernovae explode as frequently as once or 
twice every million years. The rate of explosions 
multiplied by the 
supernova energy per explosion gives  the luminosity, $ L $, of the
superbubble, which ranges from $ 10^{ 36} $ to $ 10^{ 39} $ 
ergs per second. A  superbubble lasts as long as
the  O and B stars in its cluster are  forming.
Generally,  this  lifetime is  25  to 50 million years.
The average number of supernova in a cluster lies between 
 6 to several thousand.  The luminosity of the superbubble $ L $ 
is  emitted continuously and can  be estimated as $ L \sim S
 E_{sn} $, where  $ S $ is the   
 supernova rate and $ E_{sn} $ is the energy of a  supernova.

It is found that the shells of 
superbubbles with $ L> 10^{37} \mbox{ ergs/s} $,  expand
long 
enough to break out of the disc.
Prior to breakout, these superbubble    decelerate,
as more and more mass is snow-plowed into their  shells.
However, for the stronger ones after they   breakout of the 
disc,  there is no further  surrounding  mass to snow-plow, and the 
pressure of the bubble  then accelerates the shell.
  This  situation is  unstable
to the Rayleigh-Taylor instability because   the  low density 
plasma of the superbubble  accelerates the high density shell.
As a consequence, the  part of the superbubble extending
out of the disc,
fragments into many pieces.  This is observed in the numerical
simulations of Mac Low et. al.(1989). 
 After the breakup,  the fragments continue to rise,
but eventually their  mass 
is  brought to a stop  by the galactic gravitational field,
at heights of  one or two kiloparsecs.

This does  not directly  help the dynamo because
when the superbubble mass falls back into the disc
it drags the flux back,  thus  expelling no net flux. 
  However, it is shown in section 7 that a 
  spike instability  develops
on  any  fragments that manage to hold onto
the  superbubble pressure for a long enough  time.
This spike will manage to throw mass out of the halo even if  the fragment
 falls back into the disc.

In fact,  the mass at the very top of the
spike  achieves velocities greater than  the galactic  escape velocity.
The mass at the top of the spike
 will thus drag the  flux lines 
that are  still  embedded in it, out of the galaxy.  
Each of the very short segments of these 
 lines  will   'cut' the rest of the
line, that remains  in the  disc,

The part of the shell above the midplane of the disc
will contain all the flux in the top half of the disc
and  within a sphere of radius
$ H $. The  spikes
are unstable only on the top of half of the fragments
and contain  flux and mass from one quarter of the disc.
  Similar things happen to  the part of the shell
below the midplane.

 The rest of the paper, is devoted to finding the number of cuts
in a given period of time. 
We  first estimate  the number of fragments 
into which the superbubble shell breaks. 
We assume that $ f $ of these  fragments will have a spike. Then,
after investigating the nature of the spikes, we  calculate
the number of lines of force it ejects from the
galaxy, i.e. the number of cuts it makes.

Next,    we find the total number of spikes and  cuts above the midplane 
due to a single  superbubble
Then, making use of the rate of production of superbubbles 
per unit area  and time, we calculate the number of cuts made
per square  kiloparsec per million years. 

 By considering
an annulus through the solar circle,
we determine both 
 the number of lines and the number of cuts 
in this annulus in $ t_6 $ million years.  The ratio  of these two numbers
is  the number of cuts
per line in this time. 

 From this number  we get 
 the average length of the  cut lines at any  time. From diffusion theory
with the a diffusion
coefficient $ \beta_d $ obtained from   dynamo theory,
 we  calculate the rate of spreading 
of $ \theta_d $ the  angle  of the cut lines relative to its initial direction. 
 The time it takes for this  average angle to become  larger
than $ \pi/2 $  is the dissipation time of the 
flux.  On comparing with the standard value for the dynamo growth 
time we can test  whether the dissipation time is shorter or
longer than the dynamo time of the models.

The  theoretical and observational  properties 
of superbubble can be found in  the following references. 
(Heiles, 1974, 1979, 1984, 1987, 1990,  
Weaver et al.1977, Mac Low and McCray, 1988, Mac Low et al 1989,
McCray and Kafatos 1987, Elmegreen and Clemens 1985, Kennicut et al 1989,
and Rafikov and Kulsrud 2000). 

\section{ Superbubble Properties}

For simplicity, we assume that the galactic disc
is a slab of uniform density, $ n_0 = 1.0 \mbox{ cm}^{-3} $
 and  has a half
thickness of $  H = 200 \mbox{pc} $,  (Ferri\`{e}re, 1998).
{\it This} slab contains the same interstellar mass  as  the galaxy.
We  assume that all superbubble explosions  are centered on the disc midplane.
From now on , we consider  the parts of the shell and its 
fragments only in the upper half of the disc.

For this case, the radius, $ R $,  of a single  superbubble 
before it breaks out of the disc, evolves as 
\[
R= 267 L_{38}^{1/5} t_7^{3/5} \mbox{pc} 
\]
where   $ L $, the  luminosity of the superbubble is 
$ 10^{ 38} L_{38} \mbox{ ergs/s} $ and $ t=t_7 10^{ 7} \mbox{years} $.
(Rafikov and Kulsrud, 2000 ; Mac Low and McCray, 1988;
Weaver et al. 1977.)

Differentiating this with respect to $ t $
 we find the velocity of expansion
\[ 
v = 15.7  L_{38}^{1/5} t_7^{-2/5} \mbox{km/s} 
\]

From these results we can determine the luminosity
$ L_1 $ at which a  superbubble is just able to  break
out of the disc. We use   the fact that  the superbubble  stops expanding
horizontally when its velocity drops to the interstellar velocity
of $ 10 \mbox{ km/s} $,  and set its horizontal radius 
equal to $ H $ at this time. (Ferri\`{e}re 1993)  (Under the uniform density 
approximation for  the disc,  the superbubble will expand spherically
up to breakout. Afterward, it will expand only in the vertical
direction.)  

 We find that
\[
L_1 = 1.46 \times 10^{ 37} \mbox{ ergs/s}
\]
and the breakout time for a superbubble with $ L>L_1 $  is
\[
t_{br}= 11.7  \left( \frac{L}{L_1} \right)^{-1/3}
\mbox{ Myr}
\]
where Myr denotes $ 10^{ 6} $ years.

The rate of production of superbubbles
is
\begin{equation}\label{eq:M}   
d \sigma = 0.10  \left( \frac{L}{L_1} \right)^{-2.3} \frac{d L}{L_1} 
\mbox{kpc}^{-2}\mbox{ Myr}^{-1}  
\end{equation}
where we have normalized the rate to the $ L_1 $ luminosity.
(Ferri\`{e}re 1993,  Kennicut et. al 1989, Elmegreen et al. 1985.) 

We are only interested in superbubbles with luminosities great 
enough to break out of the disc, i.e. $ L> L_1 $.
 The shells of these superbubbles  will fragment
by the Rayleigh-Taylor instability,  when  $ R> H $ and they
leave the disc. The growth rate of the instabilities
with horizontal wave number $ k $ has a maximum of
$ \gamma = \sqrt{ k g/2 } $ at $ k D = 1 $, where  $ D $
is the thickness of the shell.  ($ g \sim 10^{ -8} $ is
the vertical gravitational  field produced by its stars in the disc.)
(Ferri\`{e}re 1998)

We can obtain the thickness of the shell at  breakout
 by comparing the
pressure $ p $ in the hot core of the superbubble with
the pressure $ n_s \kappa_B T  $ in the shell.
The core pressure can be obtained from the energy
 deposited by the exploding supernovae into the core plasma during  the breakout
time,  $ (5/11) L t_{br}$.  (The factor
of $ 5/11 $ takes into account  the work done by the core pressure 
accelerating    the shell.)
Dividing this energy by the  spherical
volume of radius $ H $,  gives $ 3/2 $ times the core pressure.
The result of the  calculation is
\[
n_s = 40.0 (\ell^{2/3} \theta^{-1})
\mbox{cm}^{-3}
\]
where 
\begin{equation}
\ell  = \left( \frac{L}{L_1} \right)
\end{equation} 

The temperature  is not known, but  it is believed to lie
between 100 and 1000 degrees, so we normalize  it to 
  300 K and set $ T=300 \theta $.
The shell  sweeps up all the interstellar matter in the sphere
above the midplane, so 
\begin{equation} \label{eq:B} 
D=\frac{n_0 H}{3 n_s} = 1.66 (\ell^{-2/3} \theta )\mbox{ pc}
\end{equation}

The superbubble shell will break into fragments with radii on a  scale
of order $ \lambda_0 = 2 \pi/k =2 \pi D $,  with $ k =1/D $ the wave number
at   maximum growth rate, so  
\[
\lambda_0 = 10.4 (\ell^{-2/3} \theta)
\mbox{ pc} 
\]

Mac Low et. al.  (1989) have simulated the explosion of a 
superbubble with $ L \sim L_{38} $, and have displayed the fragmentation.
at the time of breakout. For this luminosity,    $ \lambda_0  
 =6 \mbox{ pc} $. A crude inspection of their figure 2c shows that the
radii  of their fragments at breakout  is  
between 12 and 15 pc. 
We  use this to set the radii  of the
fragments to $ 2 \xi \lambda_0 $, where $ \xi $ is of order one.

 The number of fragments per superbubble is thus 
\begin{equation} \label{eq:C}  
N_f = \left(  \frac{H}{2 \xi \lambda_0 } \right)^2 =
91.7 (\ell^{4/3} \theta^{-2} \xi^{-2})
\end{equation}

Each fragment will have a spike with probability $ f $.

\section{The Spike Instability}

Assume a thin disc at the top of     a fragment of the
superbubble  shell
with  surface density $ \sigma_0 $.
Ignoring the inertial forces,  the gravitational force on a unit area 
of  the shell is  $ \sigma_0  g $.
$ g = 10^{-8} \mbox{ cm } s^{-2} $,  the
 real galactic gravitational field.

Assume that the spike perturbation is azimuthally
symmetric about some vertical axis. 
We derive  the Lagrangian  equations for it, with 
 radial and vertical  Lagrangian displacements
$ R(r_0,t) $ and $ Z(r_0,t) $, where $ r_0 $ is the
Lagrangian radial independent variable. 

In the perturbation, consider a thin zonal 
disc generated by the  rotation of a line initially
extending from 
$ r_0 $  to  $ r_0 + \delta $.
After a time $ t $ these two end points are displaced to 
$ [r_0 + R(r_0,t), Z(r_0,t)] $ and
$ [r_0 + \delta + R(r_0,t) + \delta R'(r_0,t), Z(r_0,t)
+ \delta Z'(r_0,t) ] $, where the prime denotes a derivative
with respect to $ r_0 $.

\begin{figure}
\rotatebox{0}{ \scalebox{0.55}{ \includegraphics{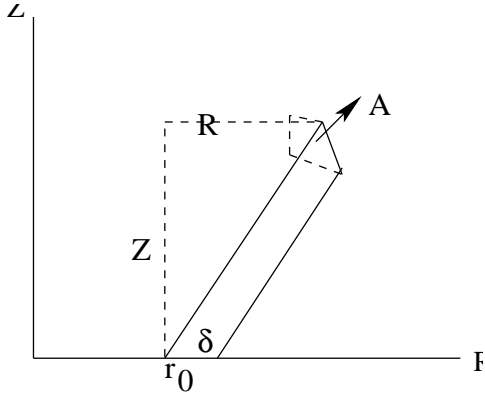} } }
\caption{The area for the spike Instability}
\label{Fig2}
\end{figure}
\vspace*{1in}

If we rotate this displaced line by  a small angle $ \epsilon $ 
about the cylindrical axis,  we will get a small vectorial  area $ {\bf  A} $ 
whose coordinates  are
\begin{eqnarray} 
A_r &=& -\epsilon \delta (r_0 +R) Z' \\  \nonumber 
A_z &=& \epsilon \delta (r_0+R)(1+R') 
\end{eqnarray} 
See figure 2.

Now $ m $,  the mass of this area,  is constant under the 
Lagrangian displacement,  
and equals $   \delta \epsilon r_0 \sigma_0 $.  The pressure force
has the components 
$ p A_r $ and $ p A_z $, with $ p = \sigma_0 g$ also a constant. 
 The gravitational force has the components 
[$ 0, - m g $].

Therefore, the (nonlinear) Lagrangian
equations of motions are
\begin{eqnarray} \label{eq:D}  
\frac{\partial^2 R}{\partial t^2 } &=& - \frac{g}{r_0}(r_0+R)Z'  \\ \nonumber 
\frac{\partial^2 Z}{\partial t^2 } &=&\frac{g}{r_0}(r_0+R)(1+R') -g
\end{eqnarray} 

Take   $ R $ and $ Z $  proportional to $ e^{\gamma t} $.  Then 
the  linear equations are
\begin{eqnarray} 
\gamma^2 R &=& -  g Z' \\ \nonumber 
\gamma^2 Z &=& g (\frac{R}{r_0} +R')
\end{eqnarray}
Setting  $ \gamma^2 = k g $, where $ k $ gives the initial spatial
dependence of the  mode, 
we find  the linear solution is
\begin{eqnarray} \label{eq:E} 
Z &=& a e^{\gamma t} J_0(k r_0) \\ \nonumber 
R &=& a e^{\gamma t} J_1(k r_0)
\end{eqnarray} 
where $ J_0, \mbox{ and} J_1 $ are Bessel functions.

When the finite thickness of the shell  $ D $ is taken into account
the growth rate is given by $ \gamma^2 = k g/(1+k^2 D^2) $
so that the maximum  growth rate occurs when $ k D =1 $, and is 
$ \gamma = \sqrt{ g/2 D } $.
Using   equation~(\ref{eq:B}) for $ D $  we find 
 the linear growth time of the spike is $ \gamma^{-1}= 1.0  \ell^{1/3} \theta^{1/2} $ 
million years.  

We need the nonlinear development of the solution  to study   the
escape of mass out of the galaxy.  

The nonlinear limit of our equations is 
\begin{eqnarray} \label{eq:F} 
\frac{\partial^2 R}{\partial t^2 } &=& - \frac{g}{r_0} R Z' \\ \nonumber   
\frac{\partial^2 Z}{\partial t^2 } &=&  \frac{g}{r_0} R R'
\end{eqnarray} 
To solve these equations,  set 
\begin{eqnarray}
R &=& \frac{\eta(r_0) }{(t_0- t)^2} \\ \nonumber 
Z &=& \frac{\nu(r_0)  }{(t_0- t)^2} \\ \nonumber 
\end{eqnarray} 

Then the nonlinear equations become
\begin{eqnarray} 
6 \eta &=& -\frac{g}{r_0} \eta \nu' \\ \nonumber 
6 \nu &=&  \frac{g}{r_0} \eta \eta'
\end{eqnarray}

Canceling $\eta$ in the first equation, we get
\[ 
\nu = C- 3 \frac{r_0^2}{g}
\]
where $ C $ is a constant to be determined. The second equation gives
\[ 
\eta^2 =\frac{6 C}{g} r_0^2 - \frac{9 }{g^2} r_0^4
\approx \frac{12 C}{g} r_0^2
\]
The linear and nonlinear solutions do not overlap.
 To  a sufficient approximation,  we can compare the two solutions
for Z at $ t=0 $, to get rough values for $ C $ and $ t_0 $.
\begin{equation}
\frac{C- 3 r_0^2/g}{t_0^2}\approx  a(1-k^2 r_0^2/4)
\end{equation}
From this we find that 
\begin{eqnarray} 
t_0^2 &=&  \frac{12}{k^2 g a} \\ \nonumber 
C &=&  a t_0^2 =  \frac{12}{k^2 g}=\frac{12 D^2}{g}
\end{eqnarray} 

Since  $ k D =1 $,  the initial radius of the spike
 is $ \sim \sqrt{ 2} D $ and  
\begin{equation} \label{eq:G} 
C=3.14 \times 10^{46} (\ell^{-4/3} \theta^2)\mbox{ cm}^2 \mbox{s} 
\end{equation} 
Note that $ C $ does not involve the initial amplitude 
$ a $.  

As $ t $ approaches $ t_0 $,  both $ Z $ and $ R $ diverge.
Before this happens, the velocity reaches  the escape velocity
and the  top of the spike  moves ballistically out
of the galaxy.
Setting the upward velocity $ \partial Z/\partial t $
at $ r_0 =0 $ and $ t=t_{esc}$, equal to $ v_{esc} 
= 400 \mbox{ km/s} $, (Binney and Tremaine 1987) we  have 
\[
\frac{\partial Z}{\partial t} =\frac{24}{k^2 g (\Delta t)^3} =v_{\mbox{esc}}.
\]
so taking $ k= 1/D $ we have
\begin{equation}\label{eq:H} 
\Delta t =  t_0 -t_{\mbox{esc}} 
= 0.37  ( \ell^{-4/9} \theta^{2/3}) \mbox{ Myr}
\end{equation} 

The height, $ h_{esc} $, at which this happens is 
\begin{equation} 
h_{\mbox{esc}} = \frac{C}{(\Delta t)^2}  
= 75 (\ell^{-4/9} \theta^{2/3}) \mbox{ pc}
\end{equation} 

At the same time the plasma at the top has spread out,
and thus,  the magnetic field originally trapped on the top
will also have spread out.   Its strength  is weakened by a power of
\begin{eqnarray}\label{eq:I}
\rho =\frac{R}{r_0}  &=& 
 \sqrt{\frac{6 C}{g}} \frac{1}{(\Delta t)^2} \\ \nonumber 
&=&  32.2 (\ell^{2/9} \theta^{-1/3})
\end{eqnarray} 
at the  time $ t_{esc} $.      

\begin{figure}
\rotatebox{0}{ \scalebox{0.55}{ \includegraphics{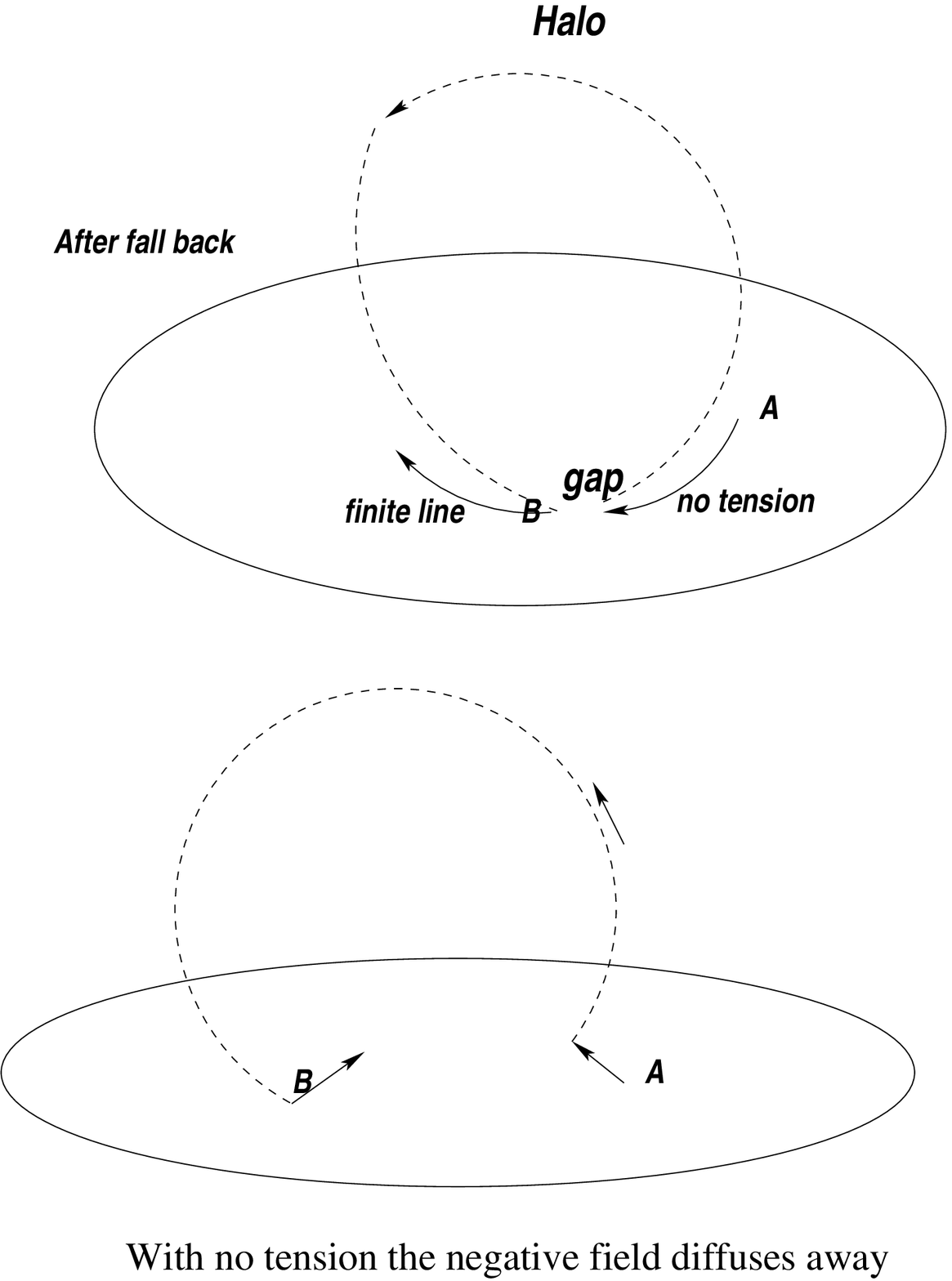}}} 
\caption{The Model for Line Cutting}
\label{Fig3}
\end{figure}
\vspace*{1in}

\section{ The Rate  of Cuts}

As a result of a single  spike, a line has been cut.  See  figure 3.
 
We will first estimate the number of cuts made by a single
spike, then the number of cuts per superbubble, and then
the rate of  cuts in a cubic kiloparsec. 

 How many cuts does a  single spike make?
 From the previous section we see that the top of the spike will be accelerated
to the escape velocity,  and leave  the galaxy, dragging  its embedded 
flux  with it.  These lines will still be connected to the
disc and each removed line will produce a cut in a disc line.

Let us take the initial horizontal diameter of the spike to be 
$ d = 2 D $.

Now, the lines in a spike will be those that come from the top half 
of its fragment, which in turn come from the top half of  
that part of the initial galactic field  above the galactic midplane.
In general, we do not know the vertical profile of the structure of the disc field 
(which is determined by the dynamo theory). But it must have the same 
profile as the initial flux in the spike. Let the initial flux in the spike  be
$ \chi B_0 (2 D H/2) $ where the profile factor $ \chi $  would be one
if the field were uniform.

  Thus, the initial number of lines
in the spike is
\begin{equation}
\phi' = \chi B_0 D H
\end{equation}

As the spike grows, the mass on the top slides downward along
its sides, and at the escape time, 
the surface density is reduced by a factor of
$ \rho^2 $ where  $ \rho $ is given by Equation~(\ref{eq:I}). 

Although the surface density  decreases by  $ \rho^2 $, the number
of field lines decreases only by one power of $ \rho $.  This is 
because in one direction,  the mass
flows along the field lines.  The lines are only removed
from the top by motions in a transverse direction to the field.
 Remember we are dealing
with extremely weak fields, so the velocities are axisymmetric and the 
fields have no affect on them.

From Equation~(\ref{eq:I}) we see that in the planer  single-spike solution, 
the spike will expand
by approximately thirty times.
It is  unlikely that the solution is valid to such a large distance,
but the thinning out of the flux and density 
about  the origin is locally correct. 
Since I am uncertain what terminates the nonlinear
solution in the horizontal direction, I conservatively 
 assume that the flux is ejected from the galaxy  
 at least  a distance $ \tau D $ from the origin, where $ \tau $ is a  factor
greater than one.  

Thus,  the  flux which is removed from the galaxy by a single spike
 and the number of resulting  cuts is 
\begin{equation}
\phi_{\mbox{spike}}= \chi \frac{\tau D H B}{\rho }
 =\chi  \tau  \frac{D}{H} \frac{H^2 B}{\rho }
=\chi 2.57 \times 10^{ -4}( \ell^{-8/9} \theta^{4/3} \tau ) \mbox{ gauss}\mbox{ cm}^2
\end{equation}    

Next, the number of cuts per superbubble is the number of cuts per 
spike times the number of fragments times the probability $ f $ that
each fragment will have a spike.  Thus,
\begin{equation}
\phi_{\mbox{sb}} = f N_f \phi_{\mbox{ spike}}
=0.0236 \times \chi  H^2 B (  \ell^{4/9} \theta^{-2/3} f \tau \xi^{-2})
\end{equation}

Now, using (Equation~(\ref{eq:M}) the  number of cuts per unit area 
 summed over all $ L $, is
\begin{eqnarray}  
\phi_{\mbox{kpc}} &=&  \int \phi_{\mbox{sb}} d \sigma \\ \nonumber 
&=& .00236 \chi  H^2  B (\theta^{-2/3}f \tau \xi^{-2}  )
 \int_1^{\infty} \ell^{4/9-2.3} d \ell \\ \nonumber 
&=& .00276  \chi  H^2 B ( \theta^{-2/3}f \tau \xi^{-2} )
 \mbox{ kpc}^{-2} \mbox{ Myr}^{-1}
\end{eqnarray} 

To get the number of cuts per unit time  of any individual  line
 in the upper quarter of the disc,  we consider
an annulus of radius 8.5 $ kpc $, thickness $ H $ and width
$ \Delta R $.  The rate of cuts in this annulus
is 
\begin{eqnarray} 
\phi_{\mbox{ann}} &=&  2 \pi R_S  \Delta R \phi_{\mbox{kpc}} \\ \nonumber 
&=& 2.76 \times 10^{ -3} \chi H^2 B  (2 \pi R_S \Delta R)
 (\theta^{-2/3} \tau f \xi^{-2}) \mbox{ Myr}^{-1}
\end{eqnarray} 

The initial number of lines 
in the top {\it quarter} of the disc and in this annulus  is

\[
\Psi = 0.5 \chi B  H \Delta R \mbox{ gauss} \mbox{ cm}^2
\]

Clearly, $  N_{\mbox{cuts}} $,  the total  of cuts which any line in the
annulus suffers 
in $ t_6 \mbox{ Myr} $  is  
\[
N_{\mbox{cuts}} =\frac{\phi_{\mbox{ann}}}{\Psi}=1.10 \times 10^{ -3} 
  (2 \pi R_S) t_6 ( \theta^{-2/3} \tau f \xi^{-2})
\] 
The profile factor, $ \chi $  has dropped out of this ratio. 

 The  average length of every  line segment after  $ t_6 \mbox{ Myr} $
is  
\begin{eqnarray}\label{eq:N} 
\mathcal{L}  =&=&   \frac{2 \pi R_S}{N_{\mbox{cuts}}} \\ \nonumber  
&=&  \frac{909}{t_6} (\theta^{2/3} f^{-1} \tau^{-1} \xi^{2}) 
\mbox{ kpc}
\end{eqnarray}

\section{The Dissipation Time for the  Flux}

We now consider how fast the cut lines can rotate in order
to remove the {\it net} flux due to them.  This rotation 
occurs in the disc and is produced by turbulent diffusion 
there. 
The   disc diffusion coefficient 
is included in all previous work.  Is has usually been estimated
from the observed properties of interstellar turbulence,  but
the magnitude  of this turbulence is  uncertain.  In the meantime
Ferri\`{e}re has carried out a more deductive derivation  of it
that is produced by the fluid motions produced by 
supernova and superbubble explosions.
(Ferri\`{e}re 1993).  Characterizing the effective turbulence
by the disc diffusion coefficient $ \beta_d $. she finds
v  $ \beta_d \approx 5 \times 10^{25}
\mbox{cm}^2 /s $.  This diffusion coefficient 
 is much smaller than the halo coefficient, discussed
in section 3, and  is the same as the beta in 
the dynamo equations (\ref{eq:A}).

Consider any given cut portion
of a  line aligned along any  axis.
  Each end will be displaced by the interstellar turbulence,
a perpendicular  amount
 whose standard deviation is $ 2 \beta t $.  Therefore, the 
 line will be rotated from an initial direction
by  an angle $ \theta_d $ whose standard deviation
is 
\begin{equation} 
<(  \theta_d )^2 > =
\int_0^t \frac{ 8 \beta t}{\mathcal{L}^2}
= 0.44 t_9^3 ( \theta^{-4/3} f^2 \tau^2 \xi^{-4}) 
\end{equation} 
where $ t_9 $ is the time in $ 10^{ 9} $ years.

   The  dissipation time, $ t_d $, is found by  setting  
$ <\theta_d^2>= (\pi/2)^2 $. This is the time 
when all the  relevant flux is diffused to zero.
\begin{equation} \label{eq:K} 
t_d =1.66 \times 10^{ 9} ( \theta^{4/9} f^{-2/3}  \tau^{-2/3}  \xi^{4/3})
 \mbox{ years} 
\end{equation}

This estimate for $ t_d $ shows that the time to dissipate
the  flux is of order of a billion years.  In deriving  it,
I make  use of certain assumptions that are quantified by the
parameters:
\begin{itemize}
\item $ f $,  the probability that a fragment will have a 
spike instability, which is equal to the probability that the fragments 
holds their superbubble 
 pressure for a time $  t_{\mbox{esc}} $, [see Equation~(\ref{eq:H})].  

\item  $ \xi  $,   the size of the fragments normalized to $ 4 \pi D $.

\item $ \theta $,  the shell temperature normalized to 300 degrees Kelvin.

\item $ \tau $, the radius  of the top of the spike at the escape time
normalized to $ 2 D $.
\end{itemize}

These parameters 
are included in Equation~(\ref{eq:K}).
Their  values are  uncertain and 
and represent its  uncertainty.
In particular, if the spike instability did not occur
at all (because  all the fragments lose  their  superbubble  pressure
 too rapidly),
 then $ f $ would be zero, Equation~(\ref{eq:K}) 
for $ t_d $
would be infinite,
and the spike  hypothesis  of this paper would not work.

For comparison let us  take an estimate of  the   dynamo time
$ t_{dynamo} $ equal to the growth time, 
from one of the  referenced papers in section
1, (Brandenburg et. al. 1992).   
 \begin{equation} \label{eq:L} 
t_{dy} = \frac{R_0^2}{11 \beta} \sim 3 \times 10^{ 8} \mbox{years} 
\end{equation} 
a time considerably  shorter than  a billion years. Thus,
if the Brandenburg et. al.  galactic dynamo is inhibited by flux-freezing
and the galactic gravitational field, it still can work
using this new process, but with a longer  growth time given
by $ t_d $. 

Although we cannot determine the values of the parameters, we
can make certain remarks.  First $ f $ is the most uncertain
parameter since the analytic details of the fragmentation are not known.
$ f $  must be less than one, but is probably bigger than one third.
The parameter $ \tau $ is definitely larger than one 
but less than $ 4 \pi \xi $ since $ 4 \pi \xi D $ 
is the radius of the fragment. It is  probably
considerably larger than one.  One might assume that  the  factors
 $ f $ and $ \tau $  cancel. 
$ \xi $ which has been determined by comparison with the actual
 simulation of Mac Low et. al.  (1989) is probably not far from 
one.  Thus, it is reasonable to believe
that  $ t_d $ is directly proportional to $ \theta $.
$ \theta $ can easily be as small as 1/3 if the shell temperature 
is $ 100 K $ or as large as three if the temperature 1000K so this
probably gives a plausible range for $ t_d $. 
Thus, one might have either $ t_d $ is comparable with $ t_{dy} $  
in which case the previous values for the dynamo time are believable,
or it may be much larger and the dynamo would be correspondingly
slower.
Some detailed numerical  simulations are probably necessary to decide which
is the case.

How does one analytically combine our  result for $ t_d $ with the alpha-omega 
differential Equation~(\ref{eq:A})?  The simplest way
is to modify these equations by adding terms $- B_r/t_d $ and
$- B_{\theta}/\tau_d  $ to the right hand sides of the two equations
when $ .5 H < |z| <H $. In addition, one must include the mass
flow expressed in terms of the flux flow from the flux-freezing
condition, and the gravitational field. 
It should be acceptable to use {\it vacuum  boundary conditions}
on the edge of the halo. This leads to a modification of the
standard  approach and should be more realistic.

The broken pieces of  flux from one growth phase will
not interfere with the next growth phase because the strength 
of their fields are 
 relatively small. This  only holds
during  amplification. When the  stronger field is 
approached, the spike instability saturates and one can
no longer neglect the residual  field in the disc even after
it is dissipated.  This will probably saturate the 
dynamo as well, since $ t_d $ would be much longer, and the boundaries
would become completely closed.  Assuming saturation at the present epoch
 these loose 
{\it finite} pieces could
be responsible for the   random fields measured by  Rand and Kulkarni (1988).

Incidentally, we can calculate the length  of the cut lines averaged
over a decay time. This is given by averaging
 Equation~(\ref{eq:N}) over  $ t_d $,
and is equal to  the
length at $ t_d $, multiplied by  a logarithmic factor 
which is  
two or  three.
This length  of only a few kpc is about the bulk distance
 cosmic rays can propagate during their  lifetime.  These 
 cut lines  remaining after saturation suggest  a natural
method for cosmic rays to escape from the disc.

\section{ Summary and Conclusions}

I have shown by a simple example that, when flux-freezing
and the gravitational field are present the
rate of expulsion of flux in previous galactic dynamo 
theories,  can be strongly inhibited
by the mass on the the outflowing flux falling back
from the halo into the disc.  A crude estimate
of this inhibition is given by Equation~(\ref{eq:BB})
which indicates that if the halo diffusion coefficient 
is much less than $ 10^{ 29} \mbox{ cm}^2/s $ the outflow
will be completely inhibited.  

However, in this case the dynamo may be saved by
another process based on a spike instability arising
from the fragmentation of those superbubble that
succeed in breaking out of the disc.  This instability
can expel a very short length of a line of force,
cutting the rest of the line in two.  If a line is cut
into short enough pieces they  can rotate horizontally
in the disc, dissipating their   net flux and
removing it from playing 
any role in the dynamo.  This turns out to be equivalent
to expelling the line as far as the dynamo is concerned.
An estimate of how many cuts occur shows that the 
relevant amount of flux can be removed in a time of
order of a billion years.  Thus, the dynamo can still  operate,
but on  this longer time scale, to grow the galactic magnetic field.
The two processes,  the alpha- omega dynamo and the flux
annihilation by cutting fields, can be combined in a systematic theory.
One merely has to add terms giving the annihilation rate, to 
dynamo equations, and assume boundary  condition the 
compatible with the inhibition of the  flux expulsion.

Incidentally, the method of dissipating flux without
removing mass may apply to other dynamos,  such as the solar dynamo,
in which
the flux is confined by flux-freezing, and one does not
want to remove  an appreciable fraction of the 
 mass of an object to produce its  dynamo. 

Although  no direct observational
evidence   is produced for this model
it may have indirect consequences of importance
such as an  explanation for  why the observed
random magnetic fields are larger than the mean
field.  Also  the cut fields may provide
an escape route for cosmic rays.  

\section{Acknowledgment} 
I should like to acknowledge my referees whose criticism
has led to a great improvement of this paper and my 
understanding of the physics underlying it.  
This work was partially supported by the U.S. Department of Energy Grants
DE-AC02-09CH11466

\section{ References}

\noindent

Baryshnikova, Iu.; Shukurov, A., Ruzmaikin, A., and  Sokoloff, D. D.
1987
Generation of large-scale magnetic fields in spiral galaxies.
   Astron. Astroph. {\bf  177}, 27.\\

\noindent
Beck, R., Brandenburg, A., Moss, D., Shukurov, A., and Sokoloff, D. 1996
 Galactic Magnetism: Recent Developments and Perspectives. 
Annu. Rev. Astron. Astroph. {\bf  34}, 155.\\

\noindent
Binney, J. and Tremaine, S. 1987 {\it Galactic Dynamics} Princeton:
Princeton Univ. Press\\

\noindent
Brandenburg, A., Donner, K.J., Moss, D.,  Shukurov, A.M. Sokoloff, D.D. 
and Tuominen, I. 1992 Dynamos in discs and halos of galaxies.
Astron. Astroph. {\bf  259}, 453.\\

\noindent
Brandenburg, A., Donner, K.J., Moss, D.,  Shukurov, A.M. Sokoloff, D.D. 
and Tuominen, I. 1993
Vertical Magnetic Fields above the Discs of Spiral Galaxies.
Astron. Astroph. {\bf  271}, 36.\\

\noindent
Dickey, J.M. and Lockman, F.J. 1990 H I in the Galaxy.
 Annu. Rev. Astron. Astroph. {\bf  28}, 215.\\

\noindent
Dobler, W, Poezd, A., and Shukurov, A. 1996
 Galactic dynamos have movable boundaries.
Astron. Astroph. {\bf  312}. 663.\\

\noindent
Donner, W., Brandenburg, A. 1990
 Generation and interpretation of galactic magnetic fields.
 Astron., Astroph. {\bf  240}, 289.\\

\noindent 
Elmegreen, B. G. and Clemens, C. 1985
 On the formation rate of galactic clusters in clouds of various masses.
ApJ.  {\bf  294}, 523.\\

\noindent 
Ferri\`{e}re, D.   1993 
Magnetic diffusion due to supernova explosions and superbubbles in the Galactic disc.
ApJ.  {\bf  409}, 248.\\

\noindent 
Ferri\`{e}re, K. 1998 
The Hot Gas Filling Factor in Our Galaxy
ApJ.  {\bf  503}, 700.\\

\noindent 
Heiles, C. 1974 
 Low-density ionized interstellar gas as revealed 
by interstellar optical and H I radio lines. 
 ApJ.   {\bf  193L}, 31.\\

\noindent 
Heiles, C. 1976 
 Low-density ionized interstellar gas as 
revealed by interstellar optical and H I radio lines.
Astron. Soc. Pacific {\bf  88R}, 607R.\\

\noindent 
Heiles, C. 1979 H I shells and supershells.
 ApJ.   {\bf  229}, 533.\\

\noindent 
Heiles, C. 1980 
 Is the intercloud medium pervasive? 
  ApJ.   {\bf  235}, 833. \\

\noindent 
Heiles, C. 1984 
 H I shells, supershells, shell-like objects, and 'worms'.
ApJS.   {\bf 55}, 585. \\

\noindent 
Heiles, C. 1987 Supernovae versus models of the interstellar
 medium and the gaseous halo.
 ApJ.   {\bf  315}, 555.\\

\noindent 
Heiles, C. 1990
 Clustered supernovae versus the gaseous disk and halo.
 ApJ.   {\bf  354}, 489.\\
 
\noindent  
Kennicut, R.C., Edgar, B.K. and Hodge, P. 1989
Properties of H II region populations in galaxies. II - 
The H II region luminosity function 761.
ApJ.  {\bf  337}.\\

\noindent 
Kulkarni, S.R. and Fich 1985
The fraction of high velocity dispersion H I in the Galaxy. 
ApJ.  {\bf  289}, 792.\\  

\noindent 
Kulsrud, R.M. 1999  A Critical Review of Galactic Dynamos. 
Annu. Rev. Astron. Astroph.
{\bf  37}, 37.\\

\noindent 
Kulsrud, R.M. 2010 The origin of our galactic magnetic field. 
Astron. Nach.  {\bf  331}, 22.\\

\noindent 
Kulsrud, R.M., Cen, R., Ostriker, J.P., Ryu, D. 1997
The Protogalactic Origin for Cosmic Magnetic Fields.
ApJ. {\bf   480}, 481. \\

\noindent 
Kulsrud, R.M. and Zweibel, E.G.  2008
On the origin of cosmic magnetic fields.
Rep. Prog. Phys. {\bf  71}, 1.\\

\noindent  
Mac Low, M-M and McCray, R. 1988 
 Superbubbles in disk galaxies. 
ApJ.   {\bf  324}, 776.\\ 

\noindent 
Mac Low, M-M,  McCray, R., and Norman, M.,1989
Superbubble blowout dynamics.
ApJ.  {\bf  337}, 141.\\

\noindent 
McCray, R., and Kafatos, M.  1987
 Supershells and propagating star formation.
ApJ.  {\bf  317}, 190.\\

\noindent 
Moss, D. and   Brandenburg, A., 1992 
 The influence of boundary conditions on the excitation of disk dynamo modes.
Astron., Astroph. 
{\bf  256}, 371.\\

\noindent 
Moss, D., Brandenburg, A., Donner, K.J., Thomsson, M. 1993,
Models for the magnetic field of M81.
 ApJ. {\bf  409}. 179. \\

\noindent 
Moss, D. and Sokoloff, D. 2013 
 Magnetic field reversals and galactic dynamos.
Geophy. Astroph. Fluid Dynamics.
{\bf  107}, 497. \\

\noindent 
Parker, E N  1971 The Generation of Magnetic Fields in 
Astrophysical Bodies. II. The Galactic Field.
ApJ,  {\bf  163}, 255.\\

\noindent 
Parker, E.N. 1992 Fast dynamos, cosmic rays, and the Galactic magnetic field.
ApJ.  {\bf  401}, 137.\\

\noindent 
Poezd, A., Shukurov, A., and Sokoloff, D. 1993 
 Global Magnetic Patterns in the Milky-Way and the Andromeda Nebula.
  MNRAS, {\bf  264}, 285.\\   

\noindent 
Rafikov, R. and Kulsrud, R.M. 2000 
Magnetic flux expulsion in powerful superbubble explosions
 and the alpha-Omega dynamo.
M.N.R.A.S.  {\bf  314}, 839.\\

\noindent 
Rand, R.J. and Kulkarni, S.R. 1989
 The local Galactic magnetic field. 

 ApJ. {\bf  343}, 760.\\

\noindent 
 Ruzmaikin, A.A., Shukurov, A.M. Sokoloff, D.D. 1988
{\it Magnetic  Fields in Galaxies}, Kluwer Academic Publishers,
Dordrecht, Boston, London.\\

\noindent 
Spitzer, Jr.,  L.  1962 {\it Physics of Fully Ionized Gases}
Interscience, New   York.\\ 

\noindent    
 Steenbeck, M., Krause, F., R\"{a}dler, K-H, 1966
 Berechnung der mittleren LORENTZ-Feldstarke 
fur ein elektrisch leitendes Medium in turbulenter,
 durch CORIOLIS-Krafte beeinflusster  Bewegung.
Z. Naturforsch {\bf  26a}, 369.\\

\noindent 
Tenorio-Tagle, G., Silich, S.A., Kunth, D., and Terlevich, E. 1999
The evolution of superbubbles and the detection of Lyalpha
 in star-forming galaxies.
MNRAS.  {\bf  309},  332.\\

\noindent 
Weaver, R, McCray, R., Castor, J., Shapiro, P., and Moore, R. 1977
Interstellar bubbles. II - Structure and evolution.
 ApJ. {\bf  218},  377.\\

\newpage    
 \huge
{\bf  Appendix}

\normalsize

\section{The Diffusion of Flux in the Presence of 
a Gravitational Field}

\vskip .2 in 

To address the problem of the motion of flux through in the 
halo, I consider a one dimensional diffusion problem
of particles in a downward magnetic field of strength $ g $
with diffusion coefficient $ \beta $.

The one dimensional  Fokker-Planck  equation
in $ z $ is
\begin{equation}
\frac{\partial f(z,t)}{\partial t} =
\frac{\partial^2}{\partial z^2} 
 \left[ \frac{
<(\Delta z)^2>}{2} f \right] -
\frac{\partial }{\partial x}  \left[ ( <\Delta z>) f  \right]  
\end{equation}

Let us start the problem at time $ t=0 $ with 
$ f(z,0) = \delta (z) $. Define $  <(\Delta z)^2/2>) =\beta  $.
 Then, with $ <\Delta z> =- g t $,  the Fokker-Planck  equation becomes,
\begin{equation}
\frac{\partial f(z,t)}{\partial t} = \beta \frac{\partial^2 f}{\partial z^2}
+ g t \frac{\partial f}{\partial t}  
\end{equation}

Next  transform this equation by setting $ z $ to $ z' =z+g t^2/2 $ 
which gives
\begin{equation}
 \frac{\partial f(z',t)}{\partial t}
 = \beta \frac{\partial^2 f}{\partial z'^2}
\end{equation}
whose solution is
\begin{eqnarray} 
f(z',z) & = &  \frac{1}{\sqrt{4 \pi \beta t }}
\exp{ -\frac{z'^2}{4 \beta t}} \\ \nonumber   
& = &  \frac{1}{\sqrt{4 \pi \beta t }}
\exp{ -\frac{(z+g t^2/2)^2}{4 \beta t}}
\end{eqnarray} 

Now at any time $ t $ the number of particles with $ z>R $ 
is
\begin{eqnarray} 
\Phi(t)& = &  \int_R^\infty  f(z,t) dz  \\ \nonumber 
& = & \frac{1}{\sqrt{ 4 \beta t}} \int_R^\infty 
\exp{ -\frac{(z+g t^2/2)^2}{4 \beta t}} d z \\ \nonumber 
& = &  \frac{1}{\sqrt{ \pi}} 
\int_{(R+g t^2/2)/\sqrt{ (4 \beta t)}}^{\infty}
e^{-u^2} du
\end{eqnarray} 

 In other words, 
\begin{equation}
\Phi(t)= \mbox{ Erfc}  \left[ (R+g t^2/2) /\sqrt{4 \beta t} \right] 
\end{equation}
where Erfc is the compliment of the error function.
This is Equation~(\ref{eq:BB}).
\end{document}